\documentclass[12pt]{iopart}

\usepackage{bm}
\usepackage{ascmac}
\usepackage[dvipdfmx]{graphicx}
\usepackage{color}
\usepackage{ifthen}
\usepackage{array}
\usepackage{longtable}

\begin{document}

\title[Vibration isolation system for power recycling mirrors of KAGRA]{Vibration isolation system with a compact damping system for power recycling mirrors of KAGRA}

\author{Y.~Akiyama$^{1}$, T.~Akutsu$^{2}$, M.~Ando$^{3}$, K.~Arai$^{4}$, Y.~Arai$^{4}$, S.~Araki$^{5}$, A.~Araya$^{6}$, N.~Aritomi$^{7}$, H.~Asada$^{8}$, Y.~Aso$^{9}$
% ...
(KAGRA Collaboration)}
\address{$^{1}$ Graduate School of Science and Engineering, Hosei University, Koganei, Tokyo 184-8584, Japan}
\address{$^{2}$ National Astronomical Observatory of Japan, Mitaka, Tokyo 181-8588, Japan}
\address{$^{3}$ Department of Physics, The University of Tokyo, Bunkyo, Tokyo 113-0033, Japan}
\address{$^{4}$ Institute for Cosmic Ray Research (ICRR), The University of Tokyo, Kashiwa, Chiba 277-8582, Japan}
\address{$^{5}$ High Energy Accelerator Research Organization, Tsukuba, Ibaraki 305-0801, Japan}
\address{$^{6}$ Earthquake Research Institute, The University of Tokyo, Bunkyo, Tokyo 113-0032, Japan}
\address{$^{7}$ Department of Physics, University of Tokyo, Bunkyo, Tokyo 113-0033, Japan}
\address{$^{8}$ Department of Mathematics and Physics, Hirosaki University, Hirosaki, Aomori 036-8561, Japan}
\address{$^{9}$ National Astronomical Observatory of Japan, Hida, Gifu 506-1205, Japan}
% ...

\ead{ayaka.shoda@nao.ac.jp}
\vspace{10pt}
\begin{indented}
\item[]July 2018
\end{indented}

\begin{abstract}
A vibration isolation system called Type-Bp system used for power recycling mirrors has been developed for KAGRA, the interferometric gravitational-wave observatory in Japan.
A suspension of the Type-Bp system passively isolates an optic from seismic vibration using three main pendulum stages equipped with two vertical vibration isolation systems.
A compact reaction mass around each of the main stages allows for achieving sufficient damping performance with a simple feedback as well as vibration isolation ratio.
Three Type-Bp systems were installed in KAGRA, and were proved to satisfy the requirements on the damping performance, and also on estimated residual displacement of the optics.

\end{abstract}

\section{Introduction}
The first detection of the gravitational waves (GWs) from the binary black hole in 2015~\cite{GW150914} opened a new era of GW astronomy.
Furthermore, the sky localization of the GW sources by the LIGO-Virgo joint observation enabled us to observe the electromagnetic counterpart, which revealed further information about the compact binary~\cite{GW170817}.
In order to improve the parameter estimation accuracy such as polarization of the GWs~\cite{Takeda2018}, it is important to construct the GW observation network with more than three detectors.
Having more detectors also has the advantage of increasing the observation rate.

KAGRA as a fourth detector is now under construction in Japan~\cite{KAGRA2018}.
It is a GW detector with 3-km scale cryogenic interferometer constructed underground.
KAGRA is expected to join the GW observation network within next few years~\cite{LVC+KAGRA2018}.
In addition, KAGRA employs the advanced features, such as construction in underground as a quiet environment, and cryogenic mirrors for thermal noise reduction.

Figure~\ref{fig:OPToverview} shows the configuration of the main optics in KAGRA.
Each mirror is suspended as a pendulum for vibration isolation.
The pendulum-type vibration isolation system isolates a mirror from the ground motion above the resonant frequencies of the pendulum. The amount of vibration isolation is further increased by using long suspensions and multiple stages.

On the other hand, the mirror swings with a large amplitude at their resonant frequencies.
It is essential for stable operation of the GW observatory to reduce the large motion of the optics at the resonant frequencies of the suspensions.
For example, quick recovery from the large disturbance due to earthquakes or control failures using the damping system can increase the duty cycle.

In this paper, we will introduce a vibration isolation system with a newly designed mechanics for the damping control for power recycling mirrors in KAGRA, which is called Type-Bp system.
This damping system is rather small, but has the sufficient performance to damp almost all the resonant modes properly.

\section{Vibration isolation systems in KAGRA}

\subsection{KAGRA suspension overview}

\begin{figure}[h]
  \begin{center} 
       \includegraphics[width=100mm]{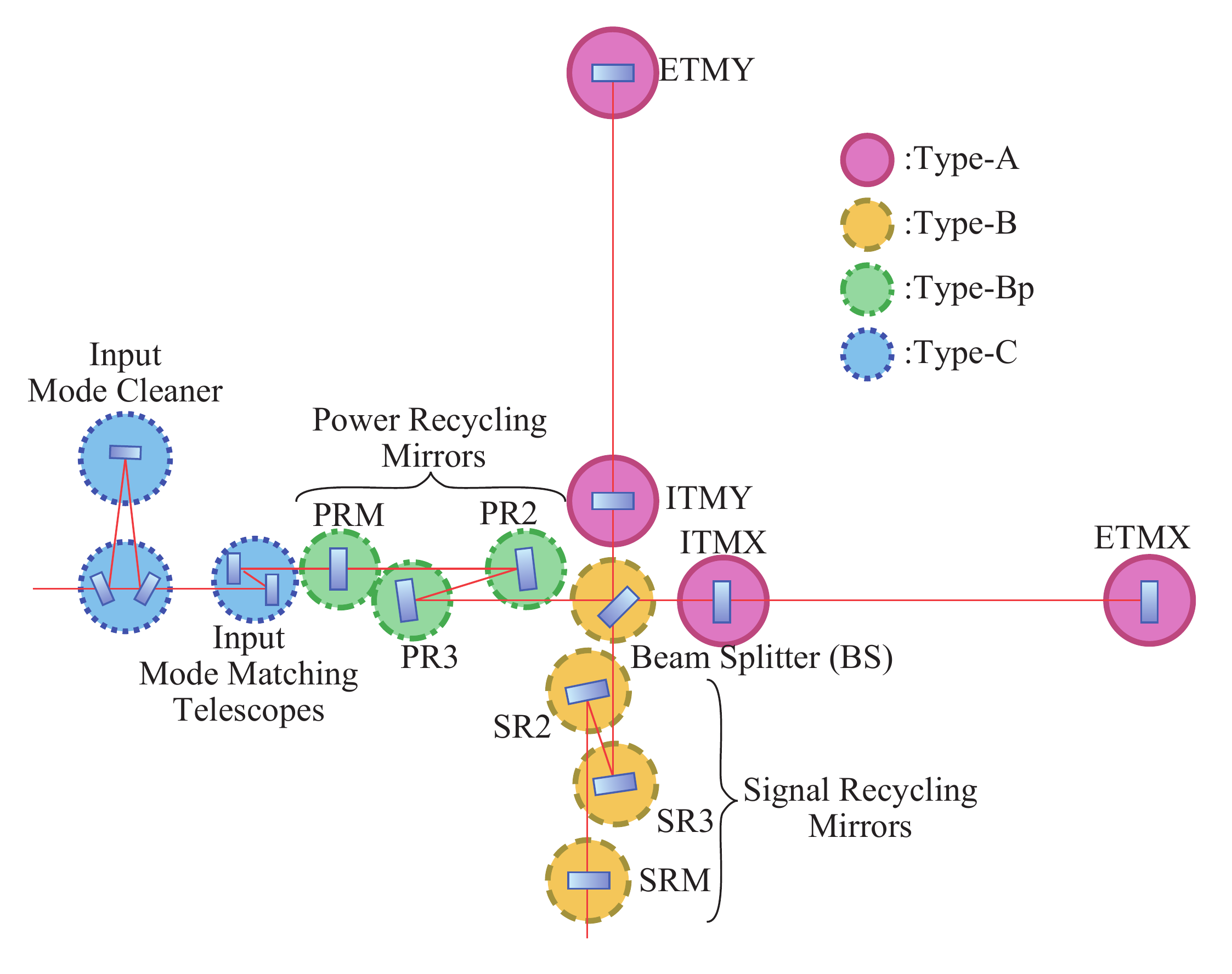}
    \caption{(Color online) Optics and vibration isolation system in KAGRA.}
    \label{fig:OPToverview}
  \end{center}
\end{figure}
\begin{figure}[h]
  \begin{center} 
       \includegraphics[width=130mm]{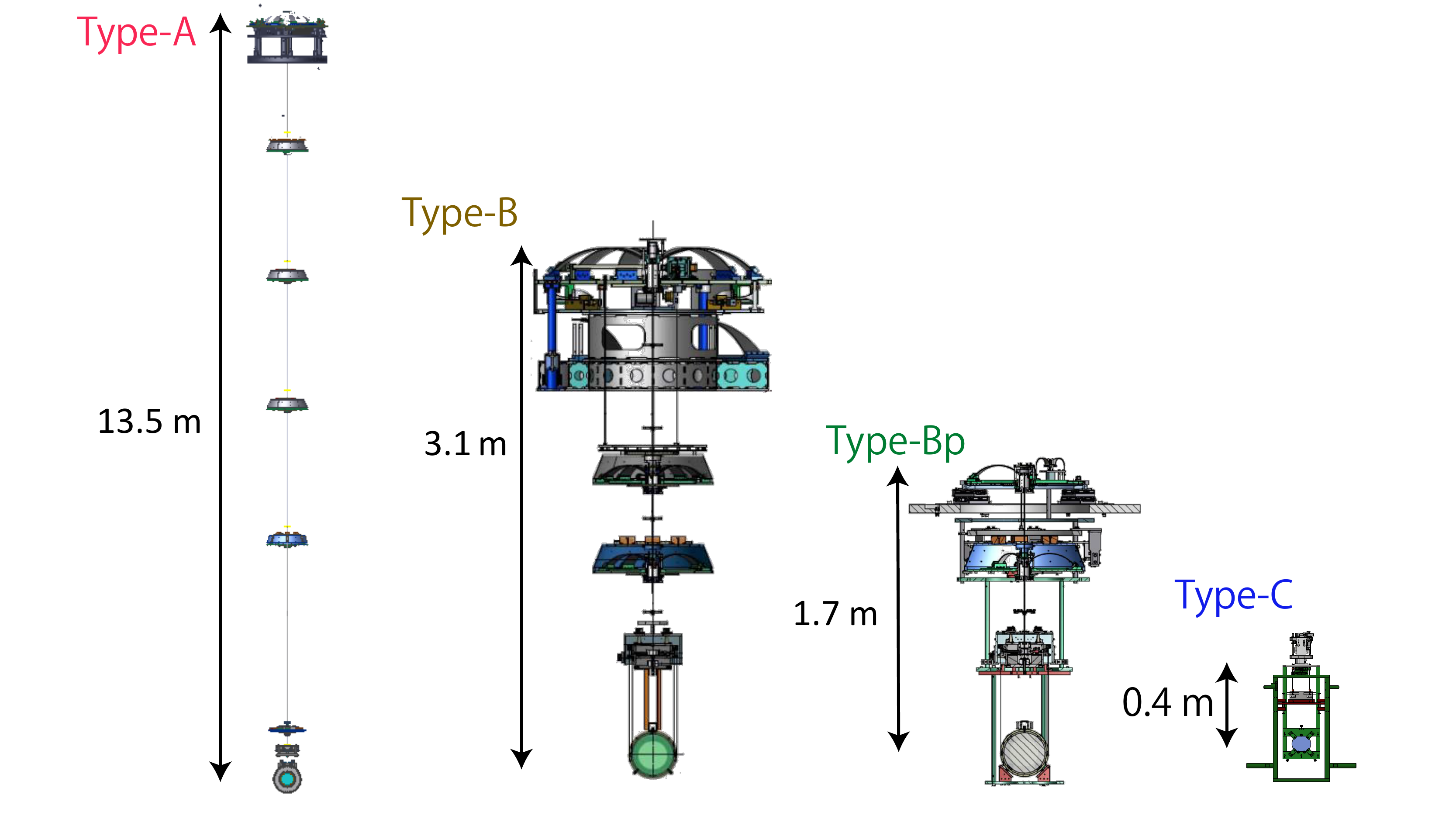}
    \caption{(Color online) Three types of vibration isolation system in KAGRA}
    \label{fig:VISoverview}
  \end{center}
\end{figure}

Circles in figure~\ref{fig:OPToverview} represent the type of the vibration isolation system used for each mirror.
We have four types of vibration systems according to their requirements on residual displacement of each mirror.
They are called Type-A, -B, -Bp, and -C systems as shown in figure~\ref{fig:VISoverview}.

A suspension of the Type-A system is used for the test masses that are most critical for the GW detection.
It has the longest tower with a height of about 13~m at room temperature, and a payload working at temperature below 20~K at the four bottom stages. 
From a pre-isolator unit composed of inverted pendulums (IPs) and a geometrical anti-spring (GAS) filter, four room-temperature stages equipped with a GAS filter inside are suspended. The IP effectively reduces the micro seismic motion that has large amplitude at around 0.3~Hz, since its resonant frequency is as low as 80~mHz. GAS filters are implemented in order to isolate the mirror from the vertical seismic motion. At the bottom, four cryogenic stages including a test mass mirror are suspended. 
Type-B is used for beam splitter (BS) and signal recycling (SR) mirrors. This suspension has a pre-isolator unit at the top similarly to the suspension of Type-A, while it has only two GAS filter stages below. 
The intermediate mass (IM) is suspended from the GAS spring, and the intermediate recoil mass (IRM) is from the base plate of the spring, independently.
And from the IM, the optics and the recoil mass (RM) are suspended independtly as well \cite{arellano2016}.
The power recycling (PR) mirrors are suspended by suspensions of the Type-Bp system. Unlike the Type-B system, they do not have the pre-isolator unit because of the constraint on the available space in vacuum chambers.
A suspension of the Type-C system, a double-stage pendulum, is used for the rather small mirrors, such as the input mode cleaners and the mode matching telescopes. Its design is the same as the one used in TAMA300~\cite{Takamori2002}.

\subsection{Requirement on Type-Bp and simulated results}
The vibration isolation system has requirements on its performance mainly in four aspects: displacement of the optic in KAGRA's observation band, $1/e$ decay time of the resonant modes, root mean square (RMS) of the velocity of the optic, and RMS of the residual angular fluctuation of the optic.
The suspension system has to damp the resonant mode enough as well as to reduce the motion of the optic due to the seismic motion and noise induced from the damping control.
The displacement of the optic has to be recovered back to the state where the interferometer can be operated soon enough even after the large disturbance caused by earthquakes or failure of interferometer control.
During the alignment of the interferometer, the orientation of the each optic has to be controlled so that the laser beam stably hits at almost the center of the optic at 3~km away.
Also, the optics have to have small enough momentum to acquire the control of the optical cavity. If an optic swings too fast, the actuators cannot capture the optic at an optical resonance point while the optic passes through.

\begin{figure}[h]
  \begin{center} 
       \includegraphics[width=40mm]{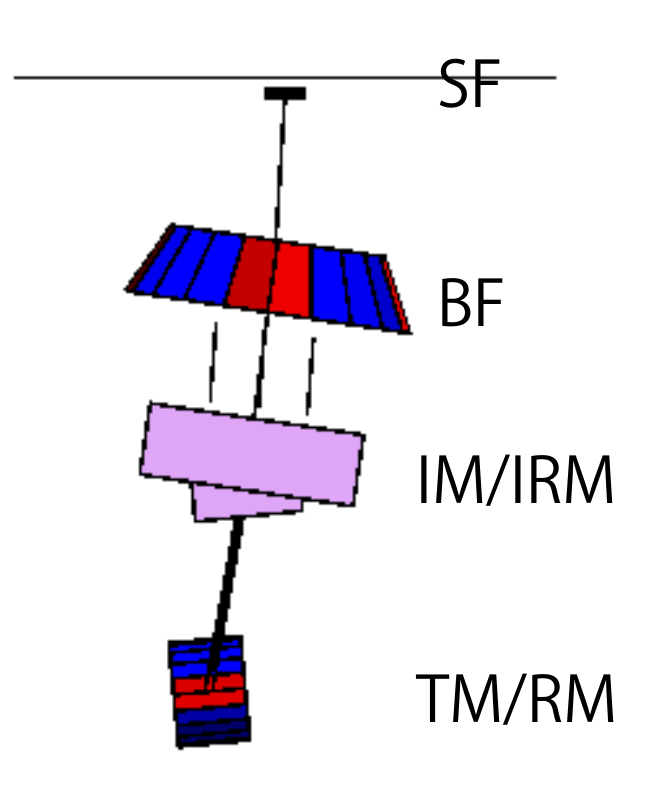}
    \caption{(Color online) Whole chain mode of the initial suspension of the Type-Bp system. This mode is difficult to damp since there is no sensors and actuators that have sufficient sensitivity and actuation efficiency.}
    \label{fig:wholechainmode}
  \end{center}
\end{figure}

Table~\ref{tab:Req} is a summary of the requirements on the PR mirrors.
Since the requirement on the displacement of the PR mirrors are about 300 times larger than that of the SR mirrors~\cite{Michimura2017, Izumi2017}, Type-Bp suspension is reasonably compact and simple.
The commissioning of the suspension control is much simpler than other suspensions since the Type-Bp suspension does not have an IP requiring careful tuning of its control filters~\cite{Sekiguchi2015}.
On the other hand, there is difficulty in reducing the damping time and angular fluctuation down below the requirement without an IP.
According to a simulation study~\cite{Sekiguchi2015}, the modes where the whole chain swings as a pendulum (see figure~\ref{fig:wholechainmode}) cannot be damped with the suspension, whereas it can be damped using the effect of its back-action on the IPs as in the case of Type-B.


\begin{table}[h]
\begin{center}
  \begin{tabular}{|l|l|l|l|}\hline
    & Requirement & \parbox{93pt}{\strut{}Simulation \\(without BFRM)\strut} & \parbox{93pt}{\strut{}Simulation \\(with BFRM)\strut} \\ \hline \hline
    Longest damping time & $6.0 \times 10$ seconds & $6.2 \times 10^2$ seconds & $2.8\times 10$ seconds\\ \hline
    RMS velocity & 6.7 $\mu$m/sec & 5.3 $\mu$m/sec & 1.4 $\mu$m/sec \\ \hline
    RMS angular fluctuation & 1.0 $\mu$rad & 1.4 $\mu$rad &  0.4 $\mu$rad \\ \hline
    \parbox{110pt}{\strut{}Mirror displacement \\at 10 Hz (with a safety factor of 10)\strut} & 5.5 $\times 10^{-15}$ m/$\sqrt{\rm Hz}$  & 2.9 $\times 10^{-15}$ m/$\sqrt{\rm Hz}$  & 3.5 $\times 10^{-15}$ m/$\sqrt{\rm Hz}$ \\ \hline
  \end{tabular}
  \caption{Requirement and simulated performance of Type-Bp vibration isolation system~\cite{Michimura2017}}
  \label{tab:Req}
  \end{center}
\end{table}

In order to solve the issue, a new compact recoil mass, BFRM, is added around so called Bottom Filter (BF), which is suspended from the ground independently from the main chain.
The sensors and actuators to monitor and actuate the relative motion between the BF and its recoil mass can damp the whole chain mode.
The damping control with the recoil mass enable the vibration isolation system to satisfy the requirements without introducing complicated mechanism nor control system that requires a lot of effort on its tuning.

\section{Design of the Type-Bp suspension}
\subsection{Mechanical design}
\begin{figure}[h]
  \begin{center} 
       \includegraphics[width=110mm]{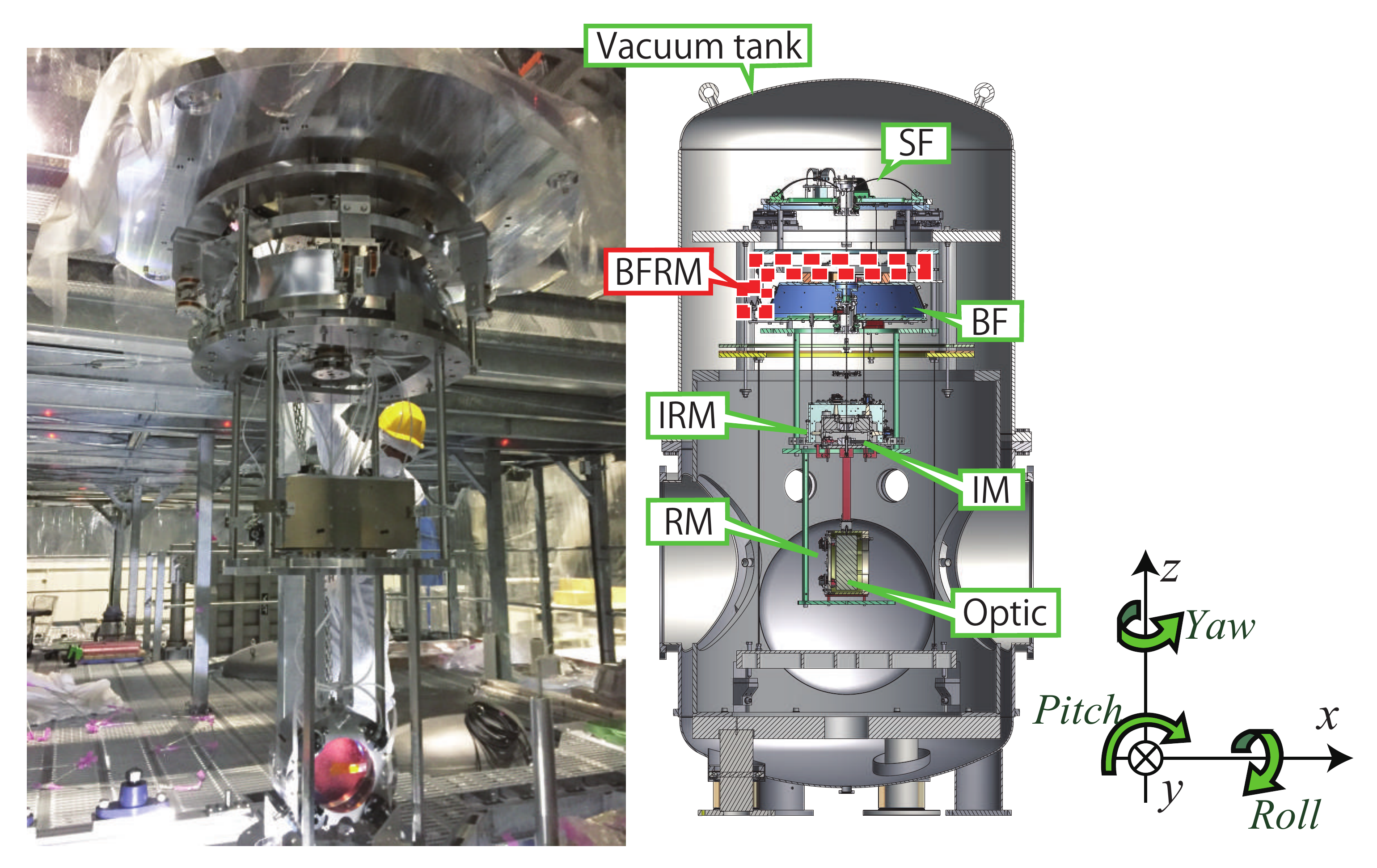}
    \caption{(Color online) A photo of the Type-Bp suspension (left) and a sectional view of the suspension inside the vacuum tank (right). The structure surrounded by the red dashed line is the BFRM, the new damping mechanism.}
    \label{fig:Type-Bp}
  \end{center}
\end{figure}

A sectional view of the Type-Bp suspension is shown in figure~\ref{fig:Type-Bp}.
We follow the definition of the axes shown in the figure throughout this paper.

\begin{figure}[h]
  \begin{center} 
       \includegraphics[width=120mm]{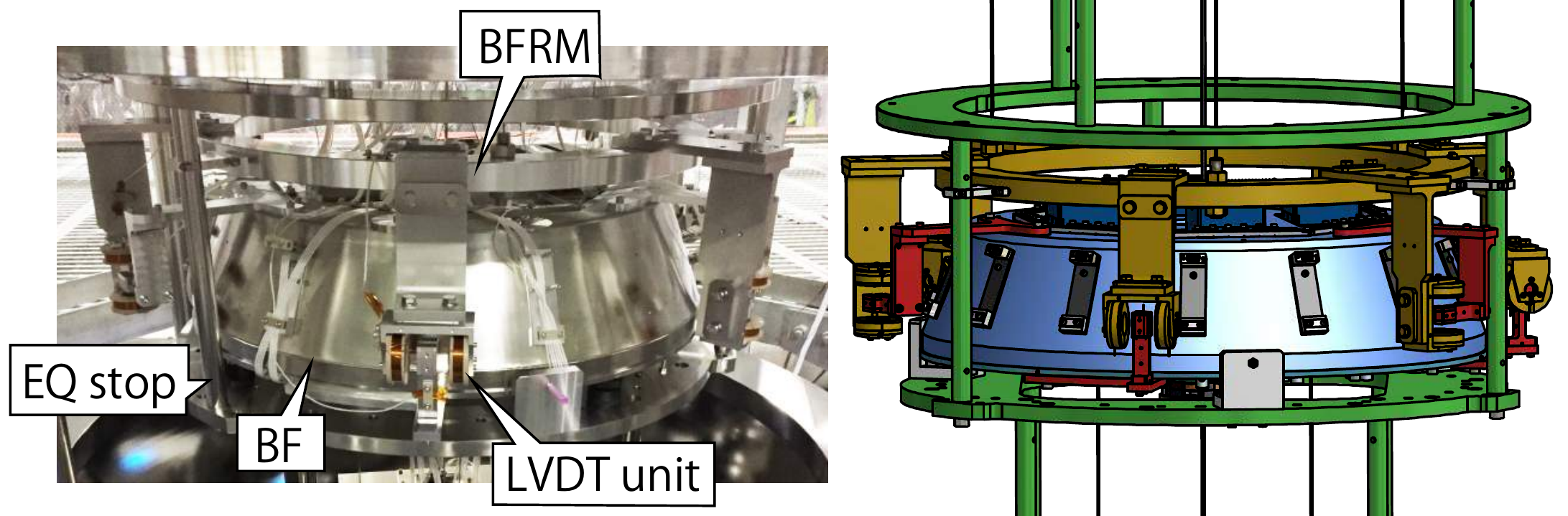}
    \caption{(Color online) A photo of the BF and newly designed BFRM (left) and the corresponding CAD drawing (right). The green, yellow, and red parts are the EQ stop, the BFRM, and the sensor and actuator units for BF damping attached on the BF, respectively.}
    \label{fig:BFRM}
  \end{center}
\end{figure}

The payload composed of the bottom two stages including optic has the same mechanism as the suspension of the Type-B system as described in \cite{Sekiguchi2015}, \cite{arellano2016} and \cite{Fujii2016}.
The payload is suspended from the BF stage that is suspended by a single maraging steel rod from another GAS filter, called the standard filter (SF).
The BFRM, the newly added recoil mass, is suspended around the BF (See figure~\ref{fig:BFRM}).
It is an aluminum ring with about 57 kg weight suspended by three maraging rods from the ground.
Each stage of the pendulum is surrounded by rigid structures called earthquake stops (EQ stops), in order to protect the suspension from serious damage caused by shocks such as one given during the installation procedure.
All the suspensions and EQ stops are suspended from the traverser stage that has a set of motors to adjust the position and the orientation of the suspension in the horizontal plane.
The traverser is attached on the rigid frame inside the vacuum chamber, and is fixed on the ground independently of the vacuum chamber. 

\subsection{Sensors and actuators}
Each stage has sensors and actuators in order to control the suspension.
The actuators are used for the interferometer controls as well as local damping control.
For the local active damping control, several types of the sensors are employed: optical lever (Oplev)~\cite{LSOplev}, an unit of a shadow sensor and a coil-magnet actuator, called Optical Sensor and Electro-Magnetic actuator (OSEM)~\cite{Carbone2012}, and two types of Linear Variable Differential Transformer (LVDT).
Each stage is activelly controlled using coil-magnet actuators attached on each stage.



\begin{figure}[h]
  \begin{center} 
       \includegraphics[width=90mm]{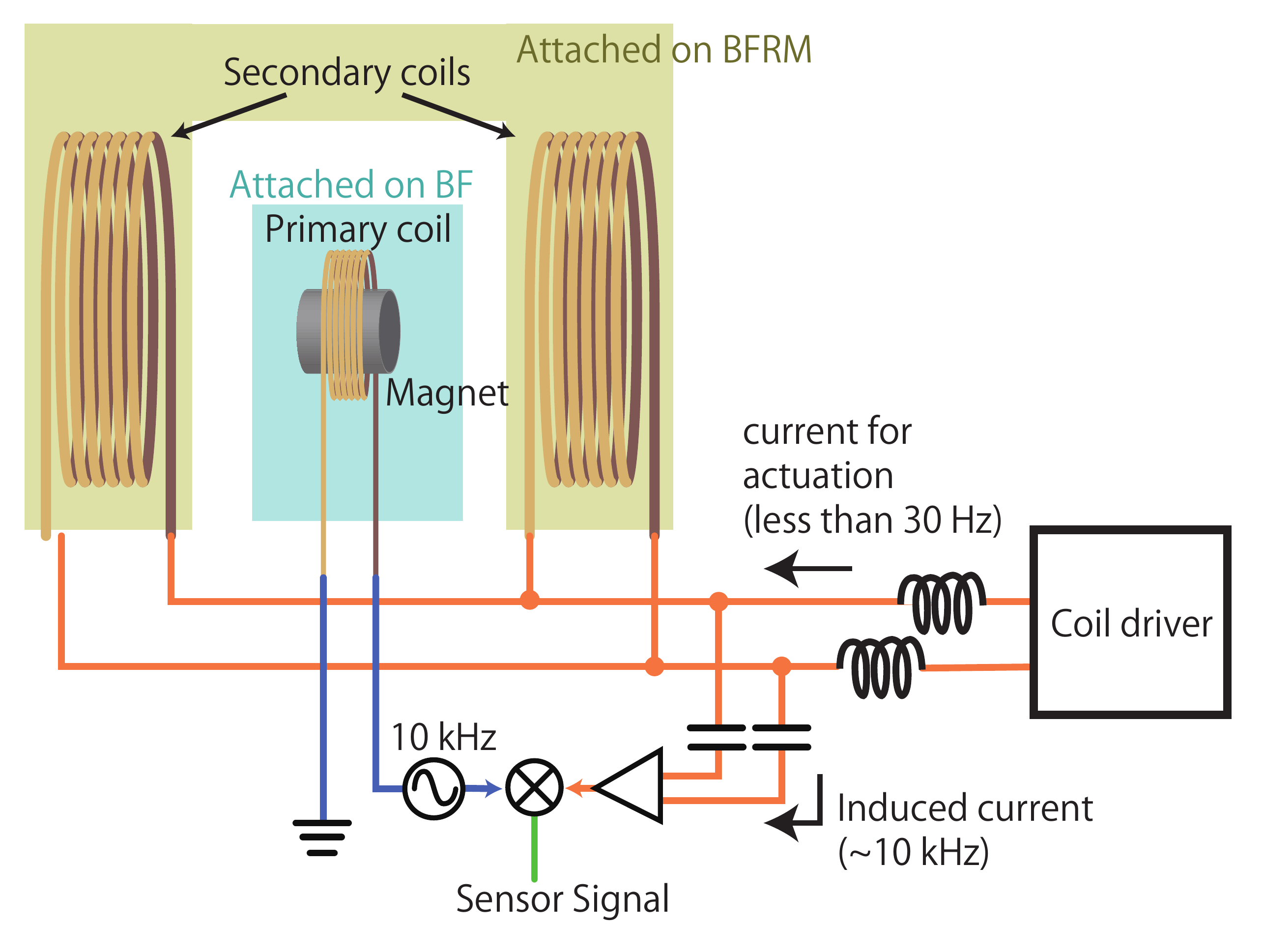}
    \caption{(Color online) The conceptual drawing of the wide-range LVDT. The primary coil that is mounted on the BF has a magnet inside, and drive the modulation signal for LVDT sensing. The secondary coils attached on BFRM play a role of the actuation coil as well as receiver coil of the induced current. The capacitors prevent the low-frequency actuation signal contaminates the sensor signal.}
    \label{fig:BFLVDT}
  \end{center}
\end{figure}

While the sensors and actuators used for the payload and the GAS filters are also same as the Type-B suspensions~\cite{Sekiguchi2015, arellano2016, Fujii2016}, the wide-range LVDTs are newly employed for the damping on the BF and the BFRM stages (See figure~\ref{fig:BFLVDT}).
The relative position between the BFRM and the BF is adjustable only by changing the height of three suspension points. 
Therefore, it is important for the sensor-actuator units to be functional even when the operating point is off from the nominal position by several millimeters as well as to have a mechanism to adjust the position of the parts attached on the BFRM.
They have one small coil as a primary coil which introduces modulated current, and two large coils as secondary coils which receive the induced electromotive force from the primary coil.
The two coils are connected so that the induced current is ideally canceled out when the primary coil is at the center of the two coils.
Also, the two secondary coils are used as actuator coils. The actuation current and the induced current are combined at the analog circuit.
They do not disturb each other in our purpose since we actuate the suspension only at frequencies lower than 30 Hz, while the induced current stays at the modulation frequencies, such as 10 kHz.

\section{Measured performance}
Three Type-Bp suspensions for PRM, PR2, and PR3 optics were assembled and installed in the KAGRA site.
They have already been utilized for the operation of a 3-km arm Michelson interferometer with a cryogenic test mass in May 2018, and have showed stable performance~\cite{bKAGRA1}.
In the following sections, measured characteristics of the PR3 are described as an example of an individual performance of the Type-Bp system.

\subsection{Damping control}
\begin{figure}[h]
  \begin{center} 
       \includegraphics[width=90mm]{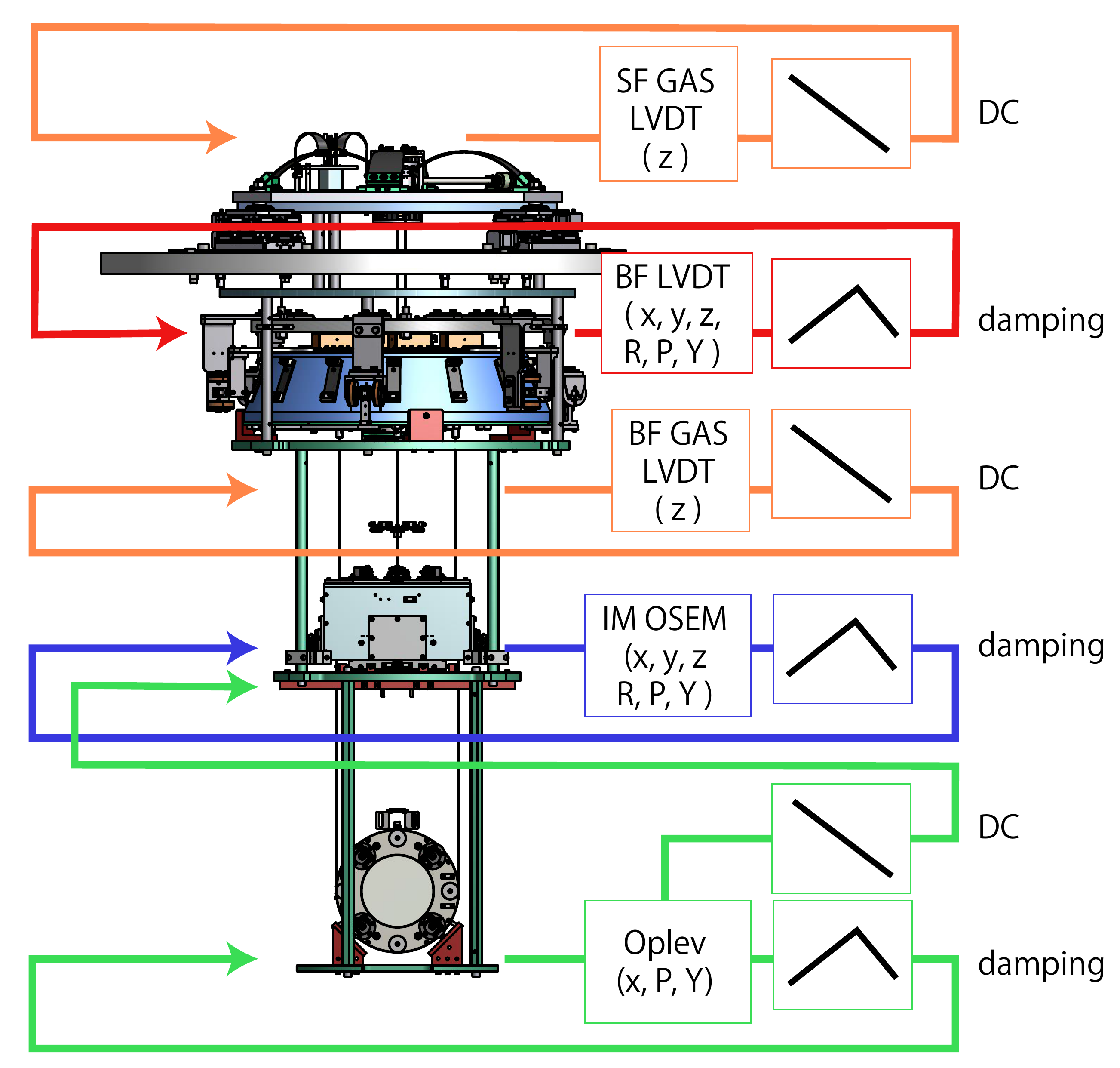}
    \caption{(Color online) Conceptual drawing of the damping control loop. Sensors attached on each stages measure the relative displacement in each DOFs between the main stage and its recoil mass, and the signals are fed back to the same stage using the coil-magnet actuators. Only Oplev loop has the hierarchical structure that signals are fed back to IM stage and optic stage. The figures in the right boxes represent the schematics of the shape of the control filters.}
    \label{fig:ControlLoops}
  \end{center}
\end{figure}

The damping control loops are engaged at all the stages for all controllable degrees of freedom as shown in figure~\ref{fig:ControlLoops}.
In order to minimize sensor noise introduced by the damping control, the damping servo filters are adjusted so that the open loop gain is larger than one only at around their resonant frequencies.
They have large control gain at the resonant frequencies by utilizing the large mechanical gain while they have low gain at higher frequencies.
Other than the damping control, stages labeled as DC in figure~\ref{fig:ControlLoops} are controlled only at frequencies lower than 0.1~Hz.
These loops are designed in order to compensate long-term drift of position and orientation of the mirror.
The feedback loops are closed in each DOFs at each stage, except for the Oplev.
The Oplev loop has the hierarchical control system in order to control the orientation of the optic.
The Oplev signal is fed back to the IM OSEMs that has larger actuation efficiency than the coil-magnet actuators at the optic stage at frequencies lower than 0.1 Hz in order to compensate drift.
The resonant peaks are suppressed by the feedback control from the Oplev to the optic similarly as other stages.

The suspensions are controlled by the digital system in the KAGRA site.
All the signals are sampled and sent to a real-time computer via the analog-to-digital converters.
After proper servo filters are applied on the signals, the feedback signals are generated at the digital-to-analog converters.

\subsection{Damping time}
\begin{figure}[h]
  \begin{center} 
       \includegraphics[width=100mm]{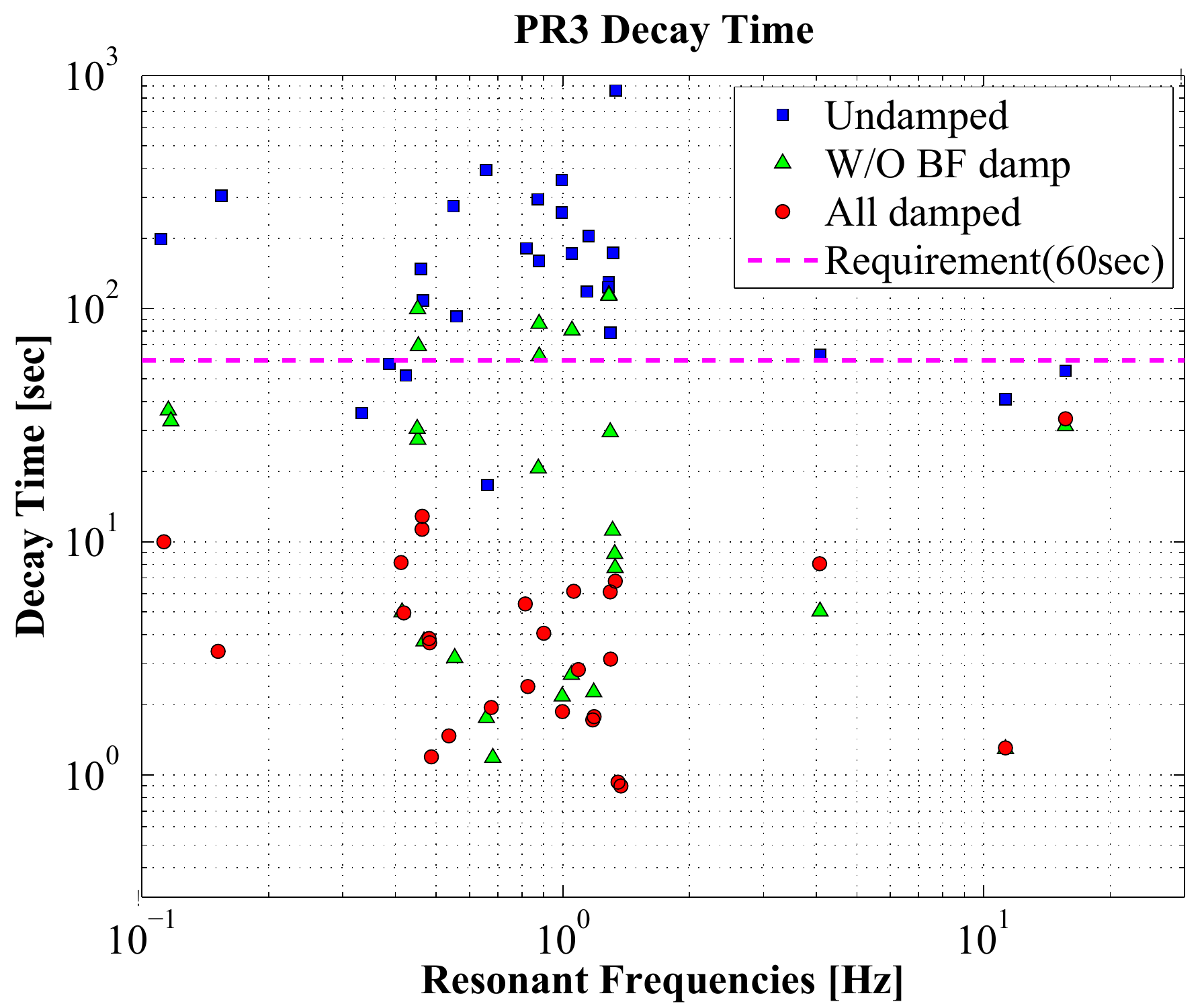}
    \caption{(Color online) Decay times of resonant modes. The blue squares, green triangles, and red circles represent the decay times when the damping controls are all off, on except for the BF stage, and all on, respectively. The dashed line shows 60 seconds, which is the requirement on the decay time.}
    \label{fig:decaytime}
  \end{center}
\end{figure}

Figure~\ref{fig:decaytime} shows how fast the amplitude of the each resonant mode decreases.
The amplitude of each mode should decay at a rate of $e^{-t/\tau_e}$, where $t$ is time, $\tau_e$ is decay time that is required for the amplitude to be decreased by $1/e$.
The Squares, triangles, and circles show the decay time $\tau_e$ without any controls, ones with damping controls except the BF/BFRM controls, and ones with all the damping loops on, respectively.
Note that a decay time of mode of differential motion of the optic and the RM in $y$-axis is not plotted in figure~\ref{fig:decaytime}.
It is only the resonant mode that cannot be observed with the local sensors.
It should not affect the interferometer operation since motion in $y$-axis less affects the interferometer alignment, and also since there is less risk to excite this mode due to rack of actuation mechanism.

It shows that some resonant modes cannot be damped within one minute without the BF damping system.
Those modes are difficult to be observed and to be actuated by the sensors and actuators other than BF LVDT actuator units.
While some of such resonant modes are motion of the BFRM that do not affect the optic motion, the whole chain mode at 0.45 Hz, for example, cannot efficiently be damped only by IM even though IM OSEMs can observe its resonance. 

\subsection{Residual velocity}
\begin{figure}[h]
  \begin{center} 
       \includegraphics[width=100mm]{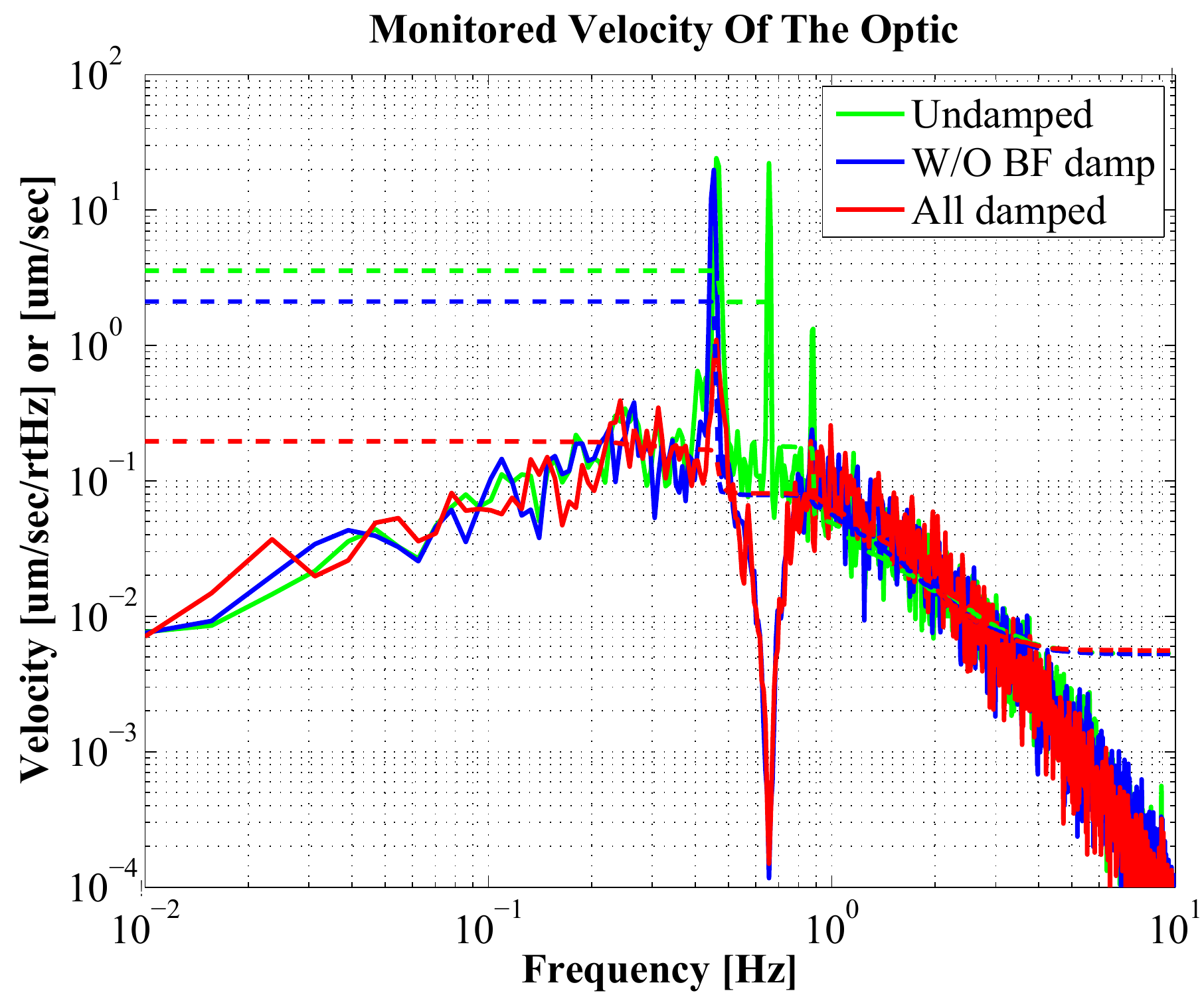}
    \caption{(Color online) Residual velocity of the optic monitored by the length-sensing Oplev. The green, blue, and red lines correspond to the velocity without any damping control, with damping on except for the BF stage, and with all damping control on, respectively. The dashed lines with each colors represent the RMS velocity accumulated below 100 Hz.}
    \label{fig:velocity}
  \end{center}
\end{figure}

Figure~\ref{fig:velocity} shows the residual velocity of the optic measured with the Oplev.
The velocity of the optic in the $x$ direction is derived by using the effect that the displacement of the optic in the Oplev signal depends on gouy phase of the beam~\cite{LSOplev}.

The green, blue, and red lines correspond to the velocity without any damping control, with damping on except for the BF stage, and with all damping control on, respectively. 
The dashed lines with each of the colors represent the RMS velocity of the optic with the corresponding damping control accumulated below 100 Hz.

Although the only displacement of the optic observed is its resonant motion due to the oplev sensing noise, the measurement still provides a useful estimate of the RMS velocity. It is known that the motion at frequencies other than the resonant frequencies, contribute in small amount according to the transfer functions from seismic motion to the optic displacement.
The total RMS of the optic velocity is 3.56~$\mu$m/sec without any damping, 2.11~$\mu$m/sec with damping control except for the BF stage, and $1.96 \times 10^{-1}$~$\mu$m/sec with all the damping on.

The resonant peak at 0.45~Hz caused by the whole chain mode is not suppressed completely even with the BF damping system, the BFRM that is about 57~kg weight is much lighter than the BF with the weight of about 100~kg.
Therefore, the BFRM follows the vibration of the main chain while some of the energy in the mode can be dissipated.

Note that dips at around 0.65 Hz are due the over-damping of the control loop on the optic stage. 
The open loop gain has large gain at 0.65 Hz that is the resonant frequency of the pendulum mode of the optic.
The dip is observed in the closed loop signal since the motion caused by the resonant mode is small compared to the open loop gain.

\subsection{Residual angular fluctuation}
\begin{figure}[h]
  \begin{center} 
       \includegraphics[width=100mm]{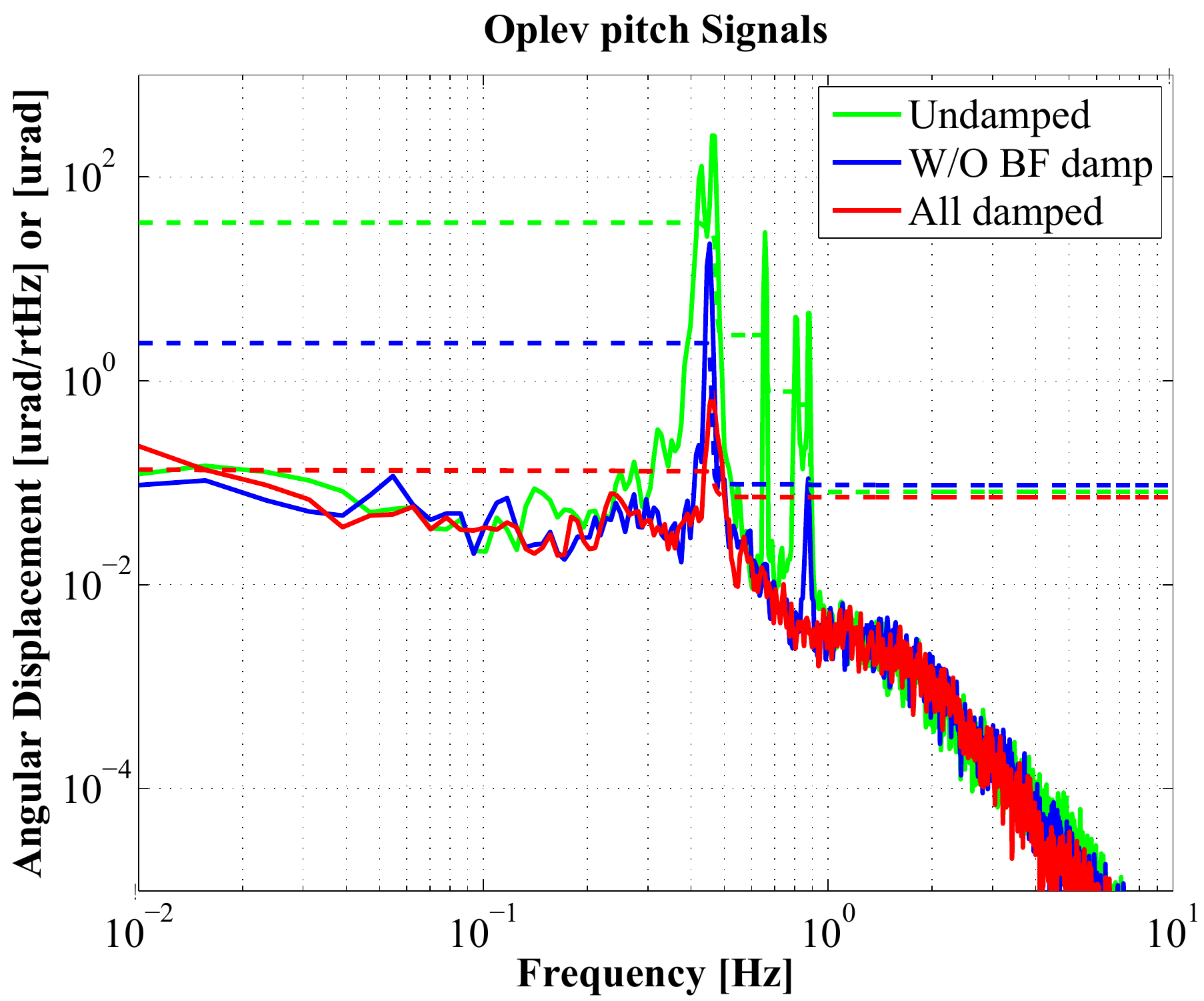}
    \caption{(Color online) Residual pitch motion of the optic monitored by the Oplev. The green, blue, and red lines correspond to the velocity without any damping control, with damping on except for the BF stage, and with all damping control on, respectively. The dashed lines with each colors represent the RMS velocity accumulated below 100 Hz.}
    \label{fig:residualPIT}
  \end{center}
\end{figure}

\begin{figure}[h]
  \begin{center} 
       \includegraphics[width=100mm]{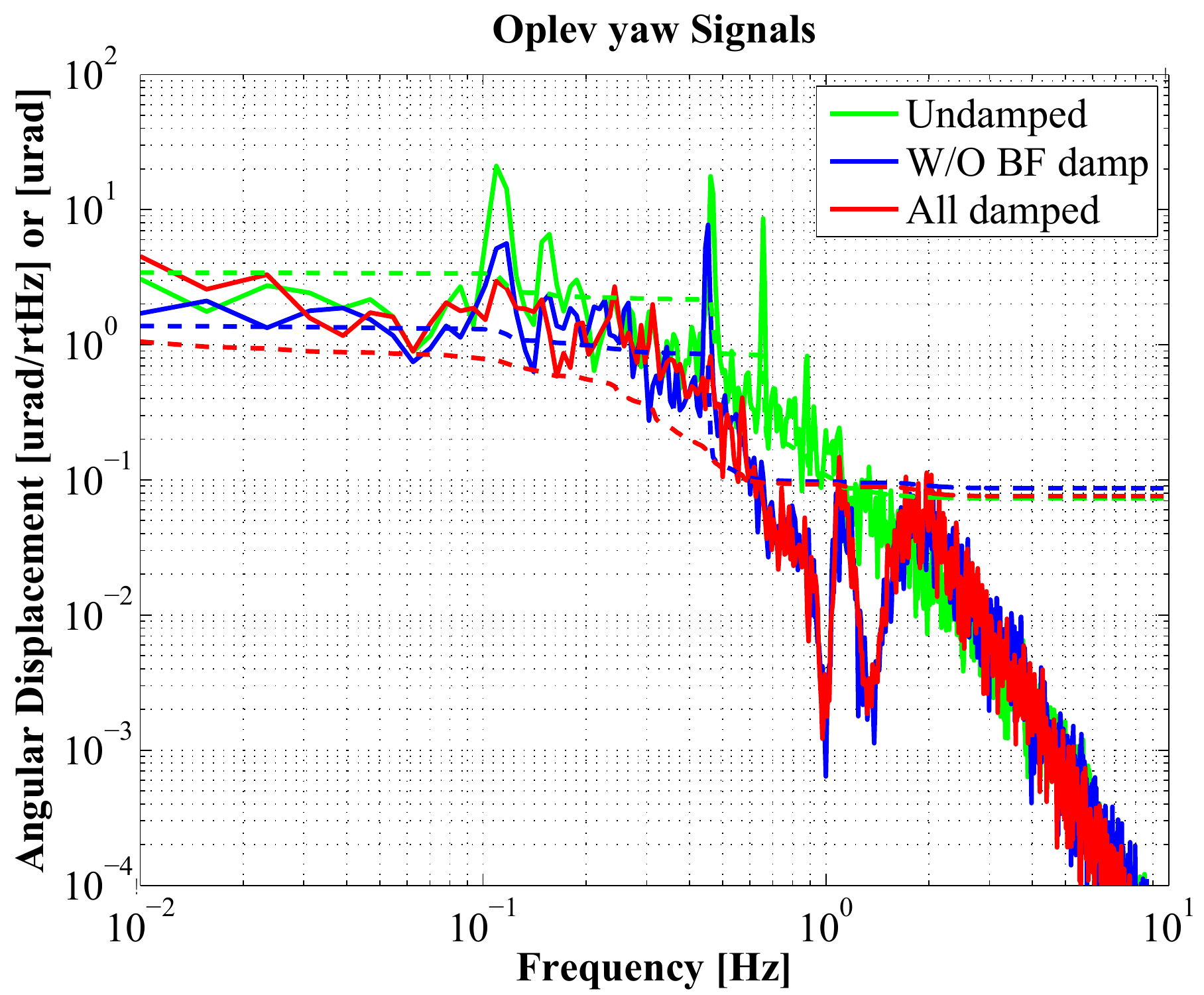}
    \caption{(Color online) Residual yaw motion of the optic monitored by the Oplev. The green, blue, and red lines correspond to the velocity without any damping control, with damping on except for the BF stage, and with all damping control on, respectively. The dashed lines with each colors represent the RMS velocity accumulated below 100 Hz.}
    \label{fig:residualYAW}
  \end{center}
\end{figure}

The angular fluctuation of the optic are shown in figures~\ref{fig:residualPIT} and \ref{fig:residualYAW}.
As well as figure~\ref{fig:velocity}, the green, blue, and red lines correspond to the angular fluctuation without any damping control, with damping on except for the BF stage, and with all damping control on, respectively. 
The dashed lines with each of the colors represent the corresponding RMS angular fluctuation below 100 Hz.

Without the newly designed BF damping system, the angular fluctuation in pitch can be suppressed from $3.56 \times 10$~$\mu$rad to 2.34~$\mu$rad, which does not satisfy the requirement, while it can be suppressed to $1.36 \times 10^{-1}$ $\mu$rad by employing the BF damping.
It shows the pitch motion caused by the whole pendulum mode at around 0.45 Hz is effectively damped by the BF damping system.

The yaw angular fluctuation that is 3.43 $\mu$rad without any damping decreased from 1.38 $\mu$rad to less than 1.11 $\mu$rad by turning the BF damping on.
The first yaw mode where the whole chain rotates without any nodes at around 0.1 Hz is suppressed efficiently by the BF damping.
Note that the RMS of the yaw fluctuation with all the damping loops on shows an upper limit of the actual residual motion, since the Oplev sensing noise is too large to observe the residual resonant peaks.

\subsection{Vibration isolation performance}
\begin{figure}[h]
  \begin{center} 
       \includegraphics[width=100mm]{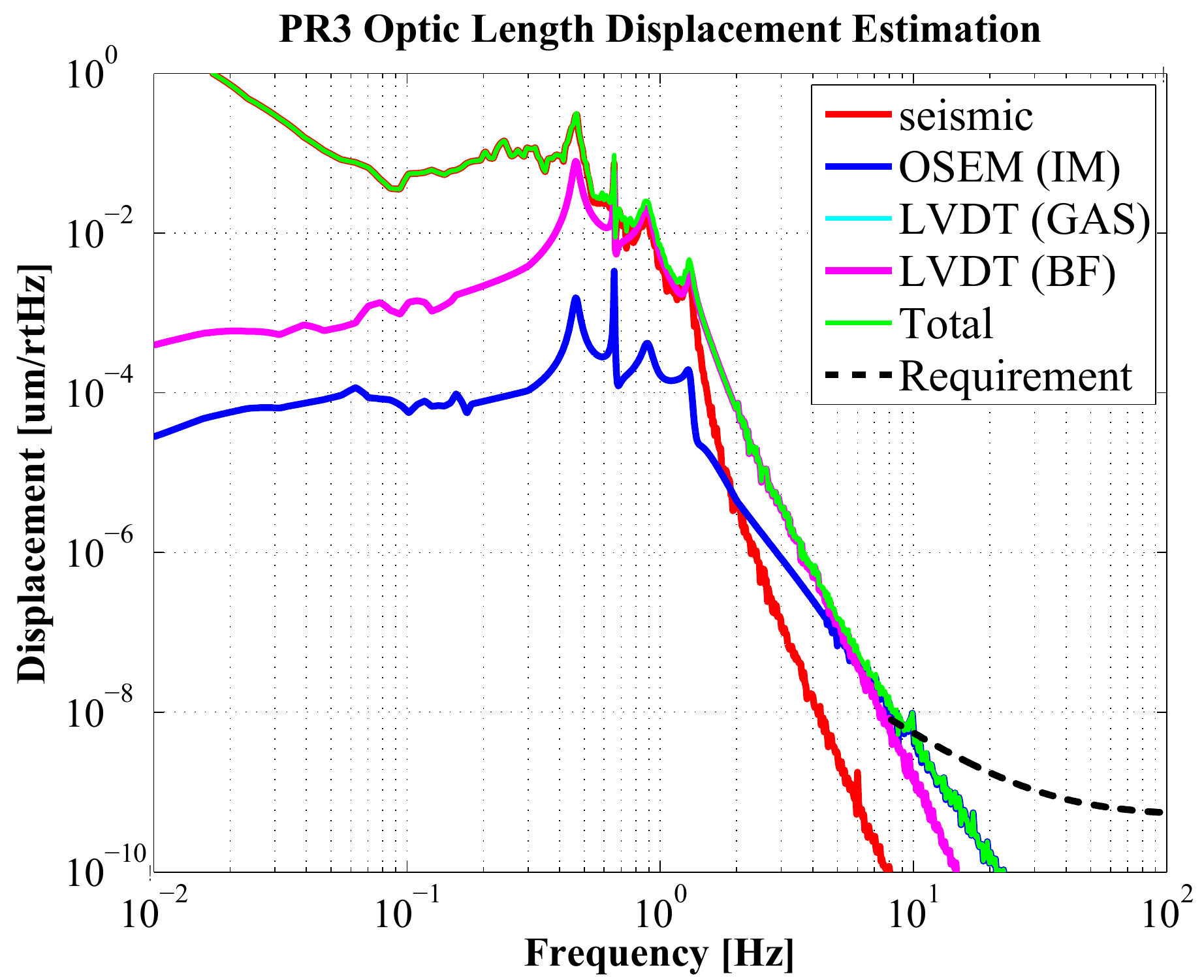}
    \caption{(Color online) Estimated fluctuation in $x$ direction of the PR3 optic when all the damping is on.}
    \label{fig:LenEst}
  \end{center}
\end{figure}

Using the rigid body model based on the measured response of the suspension, the seismic motion, and measured sensor noise, the fluctuation of the optic in $x$ direction is calculated. 
the motion of the optic at higher frequencies than the resonant frequencies cannot be measured by the local sensors due to the sensor noise and the seismic motion of the sensor.
Figure \ref{fig:LenEst} is an estimated fluctuation of the PR3 optic.
The green line is the total displacement of the PR3 optic, and the black dash line is the requirement with a safety factor of 10.
It shows that the BF LVDTs have low enough noise level for PR3. The residual displacement above 10~Hz is dominated by the IM OSEM sensor noise.

\section{Summary}
The compact vibration isolation system with proper damping control system for power recycling mirrors was realized in KAGRA.
While it has relatively compact system as tall as 1.7 m, it was designed to have sufficient vibration isolation performance and damping control system.
The new compact and simple damping system called BFRM is effective in terms of the damping control.

Three suspensions were installed as PRM, PR2, and PR3 in the KAGRA tunnel, and operated as parts of the 3-km Michelson interferomter.
Individually measured performance of the suspension is also proved to satisfy the requirements from the interferometer operation.
The $1/e$ decay time is less than 40 sec for all the resonant modes, and the residual angular fluctuation is less than 1 $\mu$rad.
Also, the residual RMS of the optic velocity is as slow as 0.2 $\mu$m/sec, such that the momentum of the optic is small enough for lock acquisition of the optical cavity.
Furthermore, the fluctuation of the optics caused by the seismic motion and sensor noise introduced by the damping control is also calculated to be small enough to achive the target sensitivity of KAGRA.

\section*{Acknowledgment}
Advanced Technology Center (ATC) in NAOJ supported the design and construction of the mechanical structures.
This work was supported by MEXT, JSPS Leading-edge Research
Infrastructure Program, JSPS Grant-in-Aid for Specially Promoted Research 26000005, MEXT Grant-in-Aid for Scientific Research on Innovative Areas 24103005, JSPS Core-to-Core Program, A. Advanced Research Networks, the joint research program of the Institute for Cosmic Ray Research, University of Tokyo, National Research Foundation (NRF) and Computing Infrastructure Project of KISTI-GSDC in Korea, the LIGO project, and the Virgo project.

\section*{References}
\bibliographystyle{unsrt}

\begin{thebibliography}{10}

\bibitem{GW150914}
B~P Abbott et~al.
\newblock {Observation of Gravitational Waves from a Binary Black Hole Merger}.
\PRL, 116(6):61102, feb 2016.

\bibitem{GW170817}
B.~P. Abbott et~al.
\newblock Gw170817: Observation of gravitational waves from a binary neutron
  star inspiral.
\PRL, 119:161101, Oct 2017.

\bibitem{Takeda2018}
H~Takeda, A~Nishizawa, Y~Michimura, K~Nagano, K~Komori, M~Ando, and K~Hayama.
\newblock Polarization test of gravitational waves from compact binary
  coalescences.
\PR {\it D}, 98:022008, Jul 2018.

\bibitem{KAGRA2018}
T~Akutsu et~al.
\newblock {Construction of KAGRA: an underground gravitational-wave
  observatory}.
\newblock {\em Progress of Theoretical and Experimental Physics},
  2018(1):013F01, 2018.

\bibitem{LVC+KAGRA2018}
B~P Abbott et~al.
\newblock {Prospects for observing and localizing gravitational-wave transients
  with Advanced LIGO, Advanced Virgo and KAGRA}.
\newblock {\em Living Reviews in Relativity}, 21(1):3, 2018.

\bibitem{arellano2016}
F~E Pe\~na Arellano et~al.
\newblock {Characterization of the room temperature payload prototype for the
  cryogenic interferometric gravitational wave detector KAGRA}.
\newblock {\em Review of Scientific Instruments}, 87(3):34501, 2016.

\bibitem{Takamori2002}
A~Takamori et~al.
\newblock {Mirror suspension system for the TAMA SAS}.
\CQG, 19(7):1615, 2002.

\bibitem{Michimura2017}
Y~Michimura et~al.
\newblock {Mirror actuation design for the interferometer control of the KAGRA
  gravitational wave telescope}.
\CQG, 34(22):225001, 2017.

\bibitem{Izumi2017}
Kiwamu Izumi and Daniel Sigg.
\newblock Advanced ligo: length sensing and control in a dual recycled
  interferometric gravitational wave antenna.
\CQG, 34(1):015001, 2017.

\bibitem{Sekiguchi2015}
T~Sekiguchi.
\newblock {\em {A Study of Low Frequency Vibration Isolation System for Large
  Scale Gravitational Wave Detectors}}.
\newblock PhD thesis, University of Tokyo, 2015.

\bibitem{Fujii2016}
Y~Fujii et~al.
\newblock {Active damping performance of the KAGRA seismic attenuation system
  prototype}.
\JPCS, 716(1):12022, 2016.

\bibitem{LSOplev}
S~Zeidler.
\newblock {\em Length-Sensing OpLevs for KAGRA}, 2017.
\newblock JGW-T1605788, Available at
  https://gwdoc.icrr.u-tokyo.ac.jp/cgi-bin/DocDB/ShowDocument?docid=5788.

\bibitem{Carbone2012}
L~Carbone et~al.
\newblock {Sensors and actuators for the Advanced LIGO mirror suspensions}.
\CQG, 29(11):115005, 2012.

\bibitem{bKAGRA1}
KAGRA collaboration.
\newblock in prep.

\end{thebibliography}

\end{document}